\documentclass[aps,twocolumn,floatfix,superscriptaddress,pra]{revtex4-1}
\usepackage{amssymb}
\usepackage{graphicx}
\usepackage{dcolumn}
\usepackage{amsmath}
\usepackage{bm}
\usepackage{float}
\usepackage{yhmath}
\usepackage{textcomp}
\usepackage{epstopdf}
\usepackage{color}
\usepackage{ulem}
\usepackage[toc,page,title,titletoc,header]{appendix}
\usepackage[colorlinks,
            linkcolor=blue,
            anchorcolor=blue,
            citecolor=blue
            ]{hyperref}
\usepackage{natbib}
\usepackage{tikz}

\begin{document}

\preprint{APS/123-QED}

\title{Excitation spectrum of a double supersolid in a trapped dipolar Bose mixture}

\author{Daniel Scheiermann}
\author{Albert Gallem\'i}
\author{Luis Santos}
\affiliation{Institut f\"ur Theoretische Physik, Leibniz Universit\"at Hannover, 30167 Hannover, Germany}

\date{\today}


\begin{abstract}

Dipolar Bose-Einstein condensates are excellent platforms for studying supersolidity, characterized by coexisting density modulation and superfluidity. The realization of dipolar mixtures opens intriguing new scenarios, most remarkably the possibility of realizing a double supersolid, composed by two interacting superfluids. We analyze the complex excitation spectrum of a miscible trapped dipolar Bose mixture, showing that it provides key insights about the double supersolid regime. We show that this regime may be readily probed experimentally by monitoring the appearance of a doublet of superfluid compressional modes, linked to the different superfluid character of each component. Additionally, the dipolar supersolid mixture exhibits a non-trivial spin nature of the dipolar rotons, the Higgs excitation, and the low-lying Goldstone modes. Interestingly, the analysis of the lowest-lying modes allows for monitoring the transition of just one of the components into the incoherent droplet regime, whereas the other remains coherent, highlighting their disparate superfluid properties.
\end{abstract}

\maketitle


\section{Introduction}


Supersolids constitute a peculiar phase of matter in which superfluidity coexists with a crystal-like density modulation~\cite{Boninsegni2012}. Although their experimental realization in helium has remained elusive~\cite{Kim2004,Balibar2010,Kim2012}, recent experiments in trapped ultracold quantum gases have explored the realization of supersolids in spin-orbit-coupled Bose-Einstein condensates~(BECs)~\cite{Li2017, Geier2021}, optical cavities~\cite{Leonard2017}, and in dipolar BECs formed by magnetic atoms~\cite{Chomaz2023, Recati2023}. In the latter case, dipolar supersolids~\cite{Tanzi2019a,Boettcher2019,Chomaz2019}, formed by ultradilute quantum droplets immersed in a superfluid halo, result from the non-trivial interplay between $s$-wave collisions, dipole-dipole interactions, quantum fluctuations~\cite{Petrov2015}, and the external confinement. 


Recent interest in dipolar supersolids has focused on their elementary excitations as a key tool for understanding their superfluid properties and associated phase transitions~\cite{Recati2023, Roccuzzo2019, Guo2019, Tanzi2019b, Natale2019, Hertkorn2019}.
Supersolids break simultaneously translational invariance, associated with the crystal order, and the $U(1)$ symmetry related to superfluidity. As a result, typical experiments performed in quasi-one-dimensional droplet arrays are expected to present two gapless Goldstone modes~\cite{Watanabe2012, Roccuzzo2019}. However, crucially, dipolar supersolids are created in the presence of an external confinement, and hence their spectrum is discretized due to finite size. For a trapped dipolar supersolid, the lowest-lying Goldstone modes are characterized, respectively, by an in-phase or out-of-phase oscillation of the condensate and the crystal structure. The in-phase mode is the center-of-mass oscillation of the whole cloud~(dipole mode), whereas the out-of-phase mode has been experimentally revealed by the absence of center-of-mass motion~\cite{Guo2019}. The two-Goldstone mode character of the spectrum has been experimentally revealed as well by the study of compressional modes, which present characteristic two-frequency excitations in the supersolid phase~\cite{Tanzi2019b, Natale2019}.

The recent realization of mixtures of two different dipolar components~\cite{Trautmann2018, Durastante2020, Politi2022} has opened new avenues for studying dipolar supersolid mixtures~\cite{Bisset2021, Bland2022b, Li2022, Arazo2023, Kirkby2023, Halder2023, Kirkby2023b}.
The crystal-like modulation may occur in the overall density in the case of miscible mixtures~\cite{Scheiermann2023}, or as the formation of alternating domains of each component (spin modulation) in immiscible ones~\cite{Bland2022b}. 
In both cases, the system enters a unique double supersolid phase where mutually interacting components remain superfluid, albeit with distinct superfluid fractions.

The double supersolid is expected to present a non-trivial excitation spectrum. 
Whereas an unmodulated binary condensate breaks a $U(1)\times U(1)$ symmetry, leading to two sound modes, a double supersolid additionally breaks translational symmetry, resulting in an extra Goldstone mode~\cite{Watanabe2012}. The three gapless modes have been recently discussed for the case of an immiscible mixture of a dipolar and a non-dipolar component in the idealized case of an infinite tube geometry, in which the mixture is not axially confined~\cite{Kirkby2023b}. However, as for the case of a single component, in experiments binary mixtures are confined in all spatial directions by a typically harmonic confinement. 
This raises the challenge of extracting information about the double-supersolid character, the distinct superfluid properties of the components, and the various phase transitions from the discrete excitation spectrum of trapped mixtures.

This paper provides a detailed analysis of the lowest-lying excitation spectrum of a trapped supersolid dipolar mixture, focusing on a miscible case. We show that even a mildly asymmetric mixture, illustrated here by the case of two components with slightly different dipole moments, results in a highly non-trivial spin nature of the dipolar rotons, the Higgs excitation, and the Goldstone modes. Moreover, in the double supersolid phase, the lowest-lying dipole and breathing modes present a characteristic three-mode structure. In particular, compressional excitations, which can be readily excited and probed in experiments~\cite{Tanzi2019b, Natale2019}, are characterized by the doubling of the superfluid breathing modes. These breathing excitations may be employed as well to reveal the markedly different superfluid fraction in each component. Finally, we discuss how the spectrum reveals the onset of the incoherent droplet regime in one component, while the other remains superfluid.

The structure of the paper is as follows. Section~\ref{sec:Model} introduces the employed model, whereas Sec.~\ref{sec:GroundStates} analyzes the different possible ground-state phases of the mixture. Section~\ref{sec:CE} evaluates and characterizes the collective excitations. In Sec.~\ref{sec:1010}, we discuss the particular case of a symmetric mixture, whereas in Sec.~\ref{sec:109unpol} we analyze the general case of an asymmetric one, in which the spectrum provides clear insights about the nature of superfluidity in the mixture. Finally, we conclude in Sec.~\ref{sec:conclusions}.


\section{Model}
\label{sec:Model}

We consider a Bose mixture of two dipolar components labeled by $\sigma=1,2$. We focus on 
magnetic dipoles with moments $\mu_{1,2}$, although similar conclusions could 
apply to electric ones. The dipoles, which are oriented along the $z$-axis by 
an external magnetic field, can belong to the same species or to 
two different ones. In addition, the atoms interact via short-range interactions 
characterized by the intra- and inter-component scattering lengths $a_{11}$, $a_{22}$, 
and $a_{12}$. The mixture is well described by the extended Gross-Pitaevskii 
equations~(eGPEs)~\cite{Smith2021,Bisset2021} 
\begin{equation}
    i\hbar \dot\psi^\sigma({\bf r}, t) = \hat H_\sigma(\mathbf r,t) \psi^\sigma({\bf r}, t),
    \label{eq:eGPEs}
\end{equation}
with 
\begin{eqnarray}
\hat H_\sigma(\mathbf r,t) &=& 
\frac{-\hbar^2\nabla^2}{2m}+
V_{\rm trap}({\bf r})+ 
\mu_{{\rm LHY}}^\sigma[n_{1,2}({\bf r},t)]
\nonumber \\
&+&\sum_{\sigma'} \int d^3r' 
U_{\sigma\sigma'}({\bf r}-{\bf r}')
|\psi^{\sigma'}({\bf r}', t)|^2, 
\label{eq:eGPE}
\end{eqnarray}
where $\psi^\sigma({\bf r}, t)$ is the condensate wave function of component 
$\sigma$, $n_\sigma({\bf r})=|\psi^\sigma({\bf r})|^2$, and 
$U_{\sigma\sigma'}(\mathbf{r})=g_{\sigma\sigma'}\delta(\mathbf{r}) + 
V_{dd}^{\sigma\sigma'}({\bf r})$, with 
$g_{\sigma\sigma'}=4\pi\hbar^2a_{\sigma\sigma'}/m$, $m$ the mass of the 
bosons~(which we assume for simplicity equal for both components), and $V^{\sigma\sigma'}_{\rm dd}({\bf r})=\frac{\mu_0\mu_\sigma \mu_{\sigma'}}{4\pi r^3} \left ( 1 - 3\cos^2\theta \right )$ the dipole-dipole interaction potential, 
where $\theta$ is the angle sustained by the polarization direction ($z$-axis) and 
${\bf r}$. The atoms are confined in a harmonic trap defined by the potential 
$V_{\rm trap}({\bf r})=\frac{1}{2}m(\omega_x^2 x^2+\omega_y^2 y^2+\omega_z^2 z^2)$. 
Quantum fluctuations are accounted for by the Lee-Huang-Yang~(LHY) term 
$\mu_{{\rm LHY}}^\sigma[n_{1,2}({\bf r},t)]=\delta E_{\rm LHY}/\delta n_\sigma$, 
where~\cite{Bisset2021}
\begin{equation}
E_{\rm LHY}=\frac{8}{15\sqrt{2\pi}}\Big(\frac{m}{4\pi\hbar^2}\Big)^{3/2}\int d\theta_k\sin\theta_k\sum_{\lambda=\pm}V_\lambda(\theta_k)^{5/2}\,
\end{equation}
is the LHY energy density, and 
\begin{equation}V_\pm(\theta_k)=\sum_{\sigma=1,2} \eta_{\sigma\sigma}n_\sigma\pm \sqrt{(\eta_{11}n_1-\eta_{22}n_2)^2+4\eta_{12}^2n_1n_2}
\end{equation}
with $\eta_{\sigma\sigma'}=g_{\sigma\sigma'}+g^d_{\sigma\sigma'}(3\cos^2\theta_k-1)$, with $g^d_{\sigma\sigma'}=\mu_0\mu_{\sigma}\mu_{\sigma'}/3$ and $\theta_k$ 
the angle sustained by momentum ${\bf k}$ with the dipole moment. 


\section{Ground states}
\label{sec:GroundStates}

A single component dipolar condensate displays three possible ground-state phases: 
an unmodulated phase with a Thomas-Fermi density profile for sufficiently strong interactions; 
a supersolid phase, characterized by a marked density modulation while still preserving superfluidity; and the incoherent droplet~(ID) regime, in which the condensate fragments into mutually incoherent droplets.



\begin{figure}[t!]
    \centering
    \includegraphics[width=\linewidth]{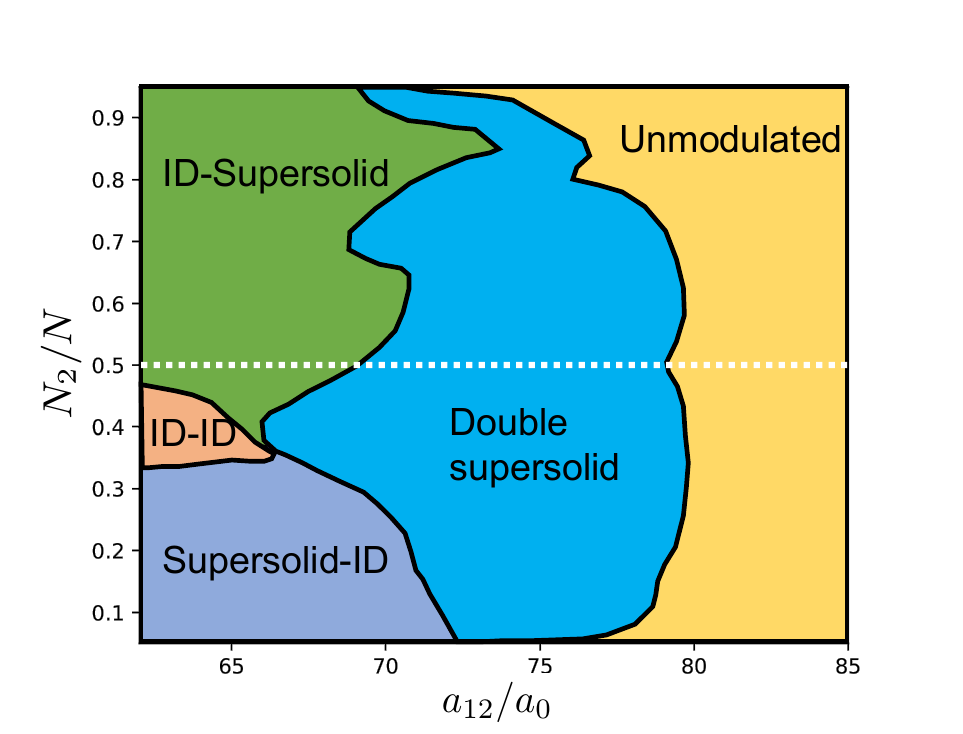}
    \caption{Ground-state phase diagram of a $^{162}$Dy mixture with $\mu_1=10\mu_B$, $\mu_2=9\mu_B$, $a_{11}=a_{22}=100\,a_0$, trap frequencies $\omega_{x,y,z}/2\pi=30$, $110$ and $90$ Hz, and a total particle number $N=3\times 10^4$. Depending on the value of $a_{12}$ and $N_2/N$, the miscible mixture may be unmodulated, double supersolid, ID-supersolid, supersolid-ID, or ID-ID. The phase boundaries are determined from the analysis of the density contrast of each component~(see text). The white dotted line indicates the case studied in Fig.~\ref{fig:3}.}
    \label{fig:1}
\end{figure}


The interplay between inter- and intra-component interactions results in a rich ground-state phase diagram for miscible mixtures~\cite{Scheiermann2023}, which we illustrate in Fig.~\ref{fig:1} for the particular case of a mixture of two $^{162}$Dy components with $\mu_1=10\,\mu_B$ and $\mu_2=9\,\mu_B$, assuming $a_{11}=a_{22}=100\,a_0$, a total number of atoms $N=3\times10^4$, and trap frequencies $\omega_{x,y,z}/2\pi=30$, $110$ and $90$ Hz~(we employ below $a_x=\sqrt{\hbar/m\omega_x}$ as the length unit). Similar phase diagrams should generally characterize other miscible mixtures. As a function of $a_{12}$ and the fraction of atoms in component $2$, $N_2/N$, the two components may be: (i) both unmodulated: this phase presents a broken $U(1)\times U(1)$ symmetry similar to that discussed in non-dipolar binary condensates; (ii) both supersolid~(double supersolid regime): in which 
due to density modulation, the system breaks (in an axially infinite system) translational symmetry, in addition to the gauge symmetry;
(iii) one component remains supersolid, whereas the other is in the ID regime~(ID-supersolid if the second component remains supersolid, or supersolid-ID in the opposite case): the system presents a broken $U(1)$ symmetry associated with the superfluid component, in addition to the broken translational symmetry; (iv) both components are in the ID regime~(ID-ID): the system behaves like a crystal of droplets, with only broken translational symmetry.

We determine the phase diagram from the density profile of each component. The contrast of the profile of component $\sigma$ is calculated from the maximal and minimal densities $n^{\rm max}_\sigma$ and $n^{\rm min}_\sigma$, as $C_\sigma=(n_\sigma^{\rm max}-n_\sigma^{\rm min})/(n_\sigma^{\rm max}+n_\sigma^{\rm min})$. Figure \ref{fig:1} shows iso-contours~(black lines) at $C_1=0.99$, and $C_2=0.1$ and $0.99$. Those lines may be employed to determine the approximate boundaries of the different regimes. 


\section{Collective excitations}
\label{sec:CE}

The different broken symmetries result in markedly different spectral properties for the different phases. In particular, for an axially-untrapped infinitely elongated mixture~(infinite tube), the unmodulated phase should present two gapless excitation branches, corresponding to the broken $U(1)\times U(1)$ symmetry, whereas the double supersolid in contrast should present three, due to the additional broken translational symmetry~\cite{Watanabe2012, Kirkby2023b}. The ID-supersolid or supersolid-ID regimes should present two branches, related to the broken translational symmetry and the remaining broken $U(1)$ symmetry associated with the component that remains superfluid. Finally the ID-ID regime should just present one branch of crystal phonons. In this paper, we focus on how these expected properties in idealized infinite tube geometries, may be revealed under typical experimental conditions, characterized by a three-dimensional confinement.


\subsection{Bogoliubov-de Gennes equations}

In order to evaluate the elementary excitations of the trapped binary mixture we linearize the eGPEs~\eqref{eq:eGPEs} around the ground states $\psi_0^\sigma({\bf r})$: 
\begin{equation}
\psi^\sigma({\bf r}, t)=e^{i\mu_\sigma t/\hbar}(\psi_0^\sigma({\bf r})+u_\sigma({\bf r})e^{i\omega t}-v_\sigma^*({\bf r})e^{-i\omega t})\,,
\end{equation}
and derive the Bogoliubov-de Gennes~(BdG) equations:
\begin{equation}
\!\!\!\begin{pmatrix}
\hat A_1 & -\hat X_{11} & \hat X_{12} & -\hat X_{12} \\
\hat X_{11} & -\hat A_1 & \hat X_{12} & -\hat X_{12} \\
\hat X_{12} & -\hat X_{12} & \hat A_2 & -\hat X_{22} \\
\hat X_{12} & -\hat X_{12} & \hat X_{22} & -\hat A_2 \\
\end{pmatrix}
\begin{pmatrix}
u_1 \\
v_1 \\
u_2 \\
v_2 \\
\end{pmatrix}
=\hbar\omega\begin{pmatrix}
u_1 \\
v_1 \\
u_2 \\
v_2 \\
\end{pmatrix}\!,
\label{eq:BdG}
\end{equation}
with $\{u_\sigma({\bf r}),v_\sigma({\bf r})\}$ the Bogoliubov eigenfunctions, 
$\hbar\omega$ the eigenenergies, 
$\hat A_\sigma ({\bf r})\equiv \hat H_\sigma({\bf r})-\tilde\mu_\sigma+\hat X_{\sigma\sigma}({\bf r})$, 
$\hat H_\sigma$ is given by Eq.~\eqref{eq:eGPE} evaluated on the ground state, $\tilde\mu_\sigma$ is the chemical potential of component $\sigma$ obtained from the ground-state calculation $\hat H_\sigma\psi_0^\sigma({\bf r})=\tilde\mu_\sigma\psi_0^\sigma({\bf r})$, and  
\begin{eqnarray}
\!\!\!\!\hat X_{\sigma\sigma'}({\bf r})\chi ({\bf r})&=&\int d{\bf r'}V^{\sigma\sigma'}_{\rm dd}({\bf r}-{\bf r}')\psi_0^\sigma({\bf r})\psi_0^{\sigma'}({\bf r'})\chi ({\bf r'})\nonumber\\
&+& \!\!\! \left ( g_{\sigma\sigma'} + \frac{\partial^2E_{\rm LHY}}{\partial n_\sigma \partial n_{\sigma'}} \right ) \!\psi_0^\sigma({\bf r})\psi_0^{\sigma'}({\bf r})\chi ({\bf r}).
\end{eqnarray}
Introducing $f_{\pm,\sigma}({\bf r})=u_\sigma({\bf r})\pm v_\sigma({\bf r})$, the 
BdG equations transform into:
\begin{align}
\begin{pmatrix}
\!
\hat B_1+2\hat X_{11}&\! 2\hat X_{12} \\
2\hat X_{12} &\! \hat B_2+2\hat X_{22}\! \\
\end{pmatrix}
\!\!
\begin{pmatrix}\!
\hat B_1&\! 0 \\
0&\! \hat B_2 \!\\
\end{pmatrix}
\!\!
\begin{pmatrix}
f_{+,1} \\
f_{+,2} \\
\end{pmatrix}
\!\!=\!(\hbar\omega)^2\!
\begin{pmatrix}
f_{+,1} \\
f_{+,2} \\
\end{pmatrix}\!,
\label{eq:BdG_f}
\end{align}
with $\hat B_\sigma({\bf r}) = \hat H_\sigma({\bf r})-\mu_\sigma$. Notice that 
$f_{-,\sigma}({\bf r})=\hat B_\sigma({\bf r}) f_{+,\sigma}({\bf r})/\hbar\omega$,  
and that we have dropped the explicit dependence on ${\bf r}$ of the operators 
and eigenfunctions in Eqs. (\ref{eq:BdG}) and (\ref{eq:BdG_f}) to keep the notation lighter. 


\subsection{Characterization of the modes}

In the following, we are interested not only in the excitation energies, but also, very especially, in the nature of the corresponding eigenmodes. We first note that due to symmetry, the modes are either even or odd under mirror symmetry. In the linear regime of validity of the BdG formalism, the density modulation of component $\sigma$ associated with a given mode is $\delta n_\sigma(\mathbf{r}) = 2\psi_0^\sigma(\mathbf{r})f_{-,\sigma}(\mathbf{r})$. The total density modulation is $\delta n(\mathbf{r})=\delta n_1(\mathbf{r})+\delta n_2(\mathbf{r})$, whereas the spin modulation, i.e. the modulation of the relative density of the two components, is $\delta s(\mathbf{r})=\delta n_1(\mathbf{r})-\delta n_2(\mathbf{r})$. We introduce $S_\pm=\int d{\bf r} \left ( \delta n(\mathbf{r})^2\pm \delta s(\mathbf{r})^2\right )$, and define $Q\equiv S_-/S_+$, which quantifies for a given mode the relative weight of the density  and the spin modulations. A mode of completely dominant density~(spin) nature is characterized by $Q=1$~($-1$). In addition, the modes satisfy the normalization $S_1+S_2=1$, with $S_{\sigma}=\int d{\bf r} f_{+,\sigma}({\bf r})\,f_{-,\sigma}({\bf r})$ the contribution of component $\sigma$ to the mode. Hence $P=S_1-S_2$ quantifies the relative weight of each component in a given mode. Note that $P=1$~($-1$) characterizes a mode of only component 1~(2).

The phase modulation in component $\sigma$ is $\theta_\sigma(\mathbf{r}) \simeq f_{+,\sigma}(\mathbf{r})/\psi_0^\sigma(\mathbf{r})$. We hence introduce the observable $\eta_\sigma=\int d{\mathbf r} |f_{+,\sigma}(\mathbf{r})| |\psi_0^\sigma(\mathbf{r})| = \int d{\mathbf r} |\theta_\sigma(\mathbf{r})||\psi_0^\sigma(\mathbf{r})|^2$, which characterizes the strength of the phase fluctuations associated with a given mode. In particular, when a component $\sigma$ transitions from the supersolid into the ID regime, a mode softens which fulfills $|f_{+,\sigma}(\mathbf{r})|\simeq |\psi_0^\sigma(\mathbf{r})|$, since for that mode $f_{+,\sigma}(\mathbf r)$ becomes also a solution of the eEGPE in the ID regime. As a result, $\eta_\sigma$ of that mode approaches $1$ when the component $\sigma$ experiences the supersolid-to-ID crossover. We define as well $\lambda_\sigma = \int d{\mathbf r} |\nabla\theta_\sigma(\mathbf{r})||\psi_0^\sigma(\mathbf{r})|^2$, which quantifies the strength of the velocity field in the high density regions. The $\sigma$ component enters the ID regime when $\lambda_\sigma$ approaches zero, since in that case 
the phase variation is localized between droplets, consistent with the incoherent nature of the ID regime.

For each mode, we can hence characterize the spin/density nature, relative contribution of each component, and phase fluctuations, using the observables $Q$, $P$, $\eta_\sigma$ and $\lambda_\sigma$. Similar observables have been recently employed in the analysis of a dipolar mixture in an infinite tube geometry~\cite{Kirkby2023b}. 

As for a single-component condensate, axial breathing modes are particularly well-suited to reveal experimentally the unmodulated-to-supersolid transition~\cite{Tanzi2019b, Natale2019}. These modes are easily excited, either by modifying the axial confinement or quenching the scattering length, and analyzed by monitoring the density profile. We thus complement our study of the Bogoliubov spectrum by analyzing the response to compressional excitations. In our analysis, we apply a perturbation proportional to $x^2$ equally to both components and monitor the normalized Fourier transform $\bar{S}(\omega)$ of $\langle x^2 \rangle (t)$, which reveals the compressional modes.


\section{Symmetric mixture}
\label{sec:1010}

As a preliminary step, we start our analysis with the discussion of the very particular and especially simple case of a symmetric mixture, with two equally populated~($N_1=N_2=N/2$) components with the same mass, $a_{11}=a_{22}=a$, and $\mu_1=\mu_2=\mu$. For a miscible mixture~(which is the case in all our calculations), $\psi_0^1(\mathbf{r})=\psi_0^2(\mathbf{r})=\psi_0(\mathbf{r})$, $\hat B_1=\hat B_2 = \hat B$ and $\hat X_{11} = \hat X_{22} = \hat X$. Hence,  Eqs.~\eqref{eq:BdG_f} exactly decouple into:
\begin{align}
\!\!\!\!\!\!(\hat B+2(\hat X+\hat X_{12}))\hat B f_{+,D} &=
(\hbar \omega)^2 f_{+,D}, \label{eq:decoupled_1} \\
\!\!\!\!\!\!(\hat B+2(\hat X-\hat X_{12}))\hat B f_{+,S} &= 
(\hbar \omega)^2 f_{+,S}, \label{eq:decoupled_2}
\end{align}
where $f_{+,D}(\mathbf{r})=f_{+,1}(\mathbf{r})+f_{+,2}(\mathbf{r})$ and $f_{+,S}(\mathbf{r})=f_{+,1}(\mathbf{r})-f_{+,2}(\mathbf{r})$, characterize, respectively, the density modulations $\delta n(\mathbf{r}) = 2\psi_0(\mathbf{r})f_{+,D}(\mathbf{r})$ and the spin modulations $\delta s(\mathbf{r}) = 2\psi_0(\mathbf{r})f_{+,S}(\mathbf{r})$. As a result, the modes acquire either a purely density character~($Q=1$) or spin character~($Q=-1$).

\begin{figure}[t!]
    \centering
\includegraphics[width=0.9\linewidth]{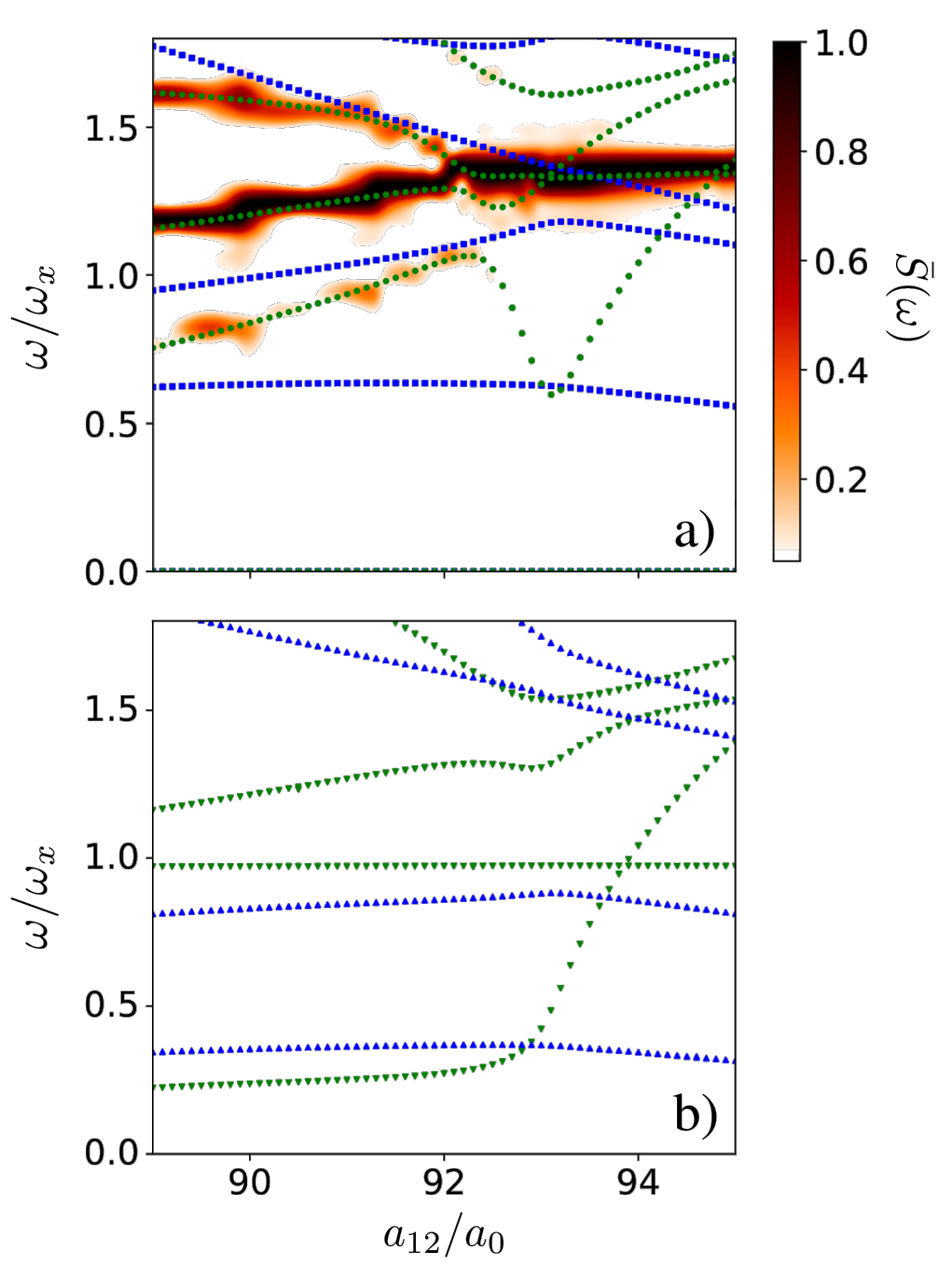}
    \caption{Modes of the symmetric mixture as a function of $a_{12}$. Panels a) and b) depict, respectively, the even and odd modes. Green~(blue) symbols indicate the density~(spin) modes. The color map in the panel a) shows the spectrum $\bar{S}(\omega)$ of excited compressional modes~(both components present the same $\bar{S}(\omega)$). Note that for each value of $a_{12}$ $\bar{S}(\omega)$ is normalized to its maximum value.}
    \label{fig:2}
\end{figure}

The ground-state total density profile $2|\psi_0(\mathbf{r})|^2$, is that of a single-component condensate with dipole moment $\mu$ and effective scattering length $a_{\mathrm{eff}}=(a+a_{12})/2$. Decreasing $a_{12}$, reduces $a_{\mathrm{eff}}$, hence increasing the relative dipolar strength~($\propto \mu^2/a_{\mathrm{eff}}$), driving eventually the unmodulated-to-supersolid transition~\cite{Scheiermann2023}. The modes across this transition are depicted in Fig.~\ref{fig:2}, where we consider the same case as in Fig.~\ref{fig:1}, but assuming $\mu_1=\mu_2=10\mu_B$, and fixing $N_1=N_2$. For this particular case, the dipolar mixture develops three central droplets at the unmodulated-to-supersolid transition.  Whereas the spin modes are mildly affected at the transition, the density modes present the characteristic softening of two degenerate roton modes, as in the case of a single-component condensate. These rotons split at the transition into a Higgs mode that abruptly hardens when entering the supersolid regime, and a Goldstone mode that softens~\cite{Guo2019, Hertkorn2019}. 

In principle, we can probe the double supersolid regime by monitoring the axial compressional modes after an $x^2$ perturbation equally applied to both components. This perturbation, however, couples only to the overall density. Hence, in a symmetric mixture, the applied compressional perturbation is orthogonal to the spin modes, 
preventing the probing of the double-supersolid nature of the spectrum. 
Similar to single-component condensates~\cite{Tanzi2019b, Natale2019}, the axial breathing density mode hybridizes with the Higgs mode and other higher-frequency modes at the unmodulated-to-supersolid transition. This leads to a hardening of the breathing mode (crystal mode) and a softening of a superfluid mode, se Fig.~\ref{fig:2}~(a). Note that, as in Ref.~\cite{Natale2019}, the additional observed even density mode, which results in a three-droplet scenario from the relative breathing of the droplets and the halo, is only weakly affected by the perturbation~(this mode is absent in the two-droplet scenario discussed in the next section).



\begin{figure*}[t!]
    \centering    
    \includegraphics[width=\linewidth]{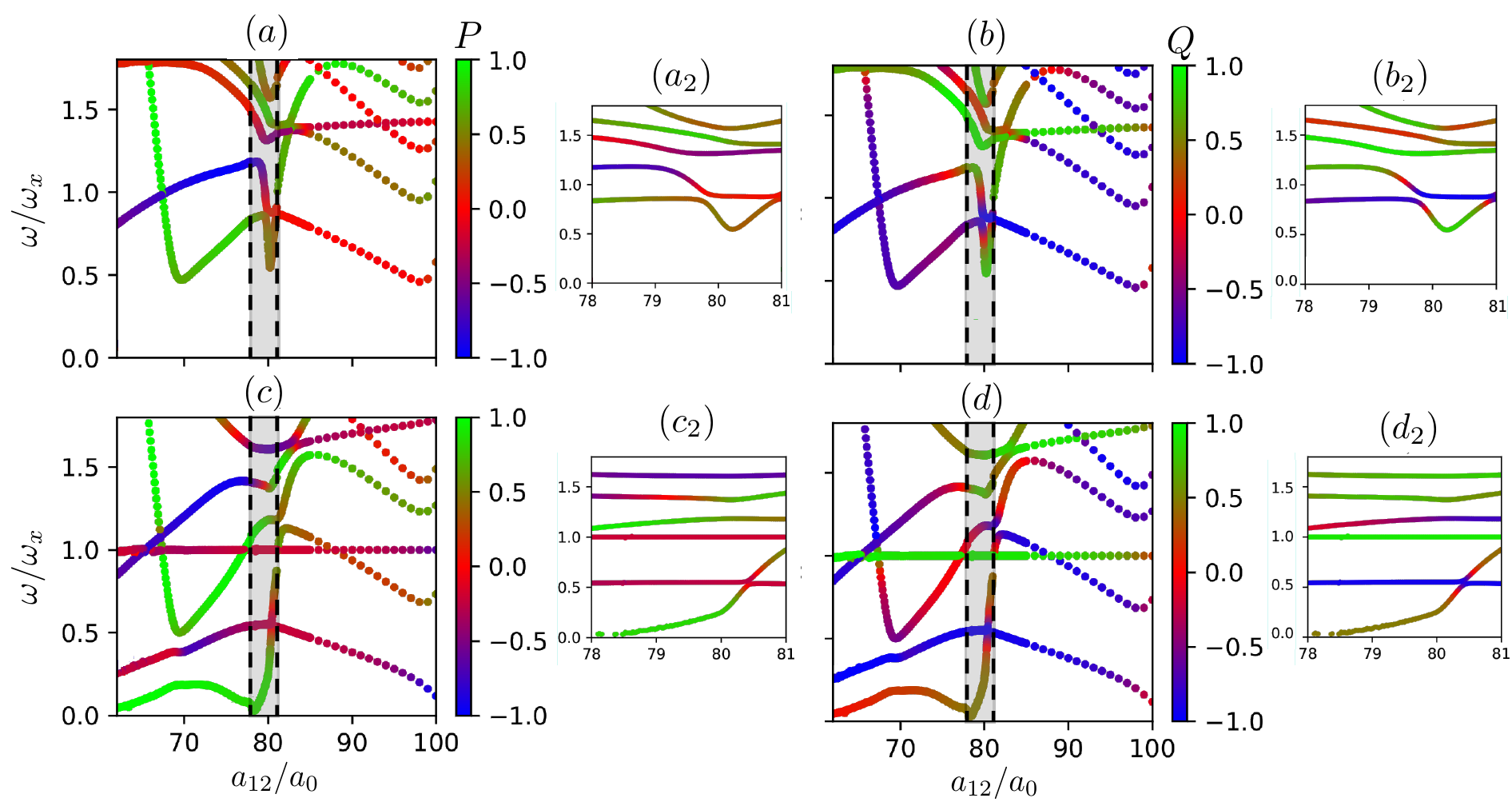}
    \caption{Lowest-lying excitations as a function of $a_{12}$ for the mixture considered in Fig.~\ref{fig:1}, assuming $N_{1,2}=N/2$. Even-parity modes are depicted in panels (a) and (b), whereas odd modes are plotted in panels (c) and (d). The color scale in panels (a) and (c) displays $P$, which characterizes the relative weight of each component, with $P=1$~($P=-1$) meaning a mode purely of component 1~(2). The color scale in panels (b) and (d) denotes $Q$, which characterizes the density versus spin character, with $Q=1$~($Q=-1$) meaning a pure density~(spin) character. Figures (a$_2$), (b$_2$), (c$_2$), and (d$_2$) depict the same results of Figs. (a-d), but zooming into the unmodulated-to-double supersolid transition region~(shaded region in Figs.~(a--d)).}
    \label{fig:3}
\end{figure*}


\section{Asymmetric mixture}
\label{sec:109unpol}

In the special case of a symmetric mixture, density and spin modes decouple, and density excitations behave as those of an effective single-component condensate. The situation becomes substantially more complex in asymmetric mixtures, where the masses and/or interactions between the components are different. In that case $\psi^1(\mathbf{r})\neq \psi^2(\mathbf{r})$ and, crucially, spin and density features hybridize, resulting in a much richer physics. We consider below the same mixture discussed above, but with different $\mu_1=10\mu_B$ and $\mu_2=9\mu_B$, as in Fig.~\ref{fig:1}. Nevertheless, the results are to a large extend representative of other asymmetric mixtures, although the particular details of the spectrum may differ. 

Figure~\ref{fig:3} summarizes our results of the low-lying spectrum. As discussed above, we characterize the modes by determining the relative weight of the two components, given by $P$~(panels a and c) and the density versus spin character, given by $Q$~(panels b and d). The upper~(lower) panels corresponds to even- (odd-) parity modes. In contrast to the symmetric case discussed above, the mixture transitions into a two-droplet ground-state when entering the double supersolid regime. Similar to the experiment in Ref.~\cite{Tanzi2019a}, this simple two-droplet scenario significantly simplifies the analysis of the excitation spectrum.

\subsection{Unmodulated regime}

In the unmodulated regime, the two Goldstone modes characteristic of the infinite tube geometry result in the trapped case in two families of low-lying modes that exhibit an approximately decoupled density~(spin) character, corresponding to in-phase~(out-of-phase) oscillations of the two components. Note that the lowest excitation corresponds to an anti-symmetric spin mode, which eventually softens at $a_{12}\simeq a=100\,a_0$, 
leading to a phonon-like immiscibility instability and 
phase separation of the two components~\cite{Ticknor2013}. 

Similar to a single-component dipolar condensate \cite{Hertkorn2019}, the unmodulated-to-double-supersolid transition is marked by the softening of two degenerate roton modes with distinct spin character. 
As seen in Figs.~\ref{fig:3}~(b) and~(d), the even roton is dominantly a mode of the total density, whereas the odd roton presents a marked spin-density hybridization. Both rotons are dominantly contributed by the most dipolar component, i.e. component 1, see Figs.~\ref{fig:3}~(a) and (c). This behavior of the rotons stems from the catalyzation effect discussed in Ref.~\cite{Scheiermann2023}. Although the roton is dominated by the most dipolar component, the associated instability results eventually in a density modulation, which is different for both components, as a result of the strong spin-density hybridization of the odd roton mode.



\begin{figure}[t!]
    \centering
\includegraphics[width=\linewidth]{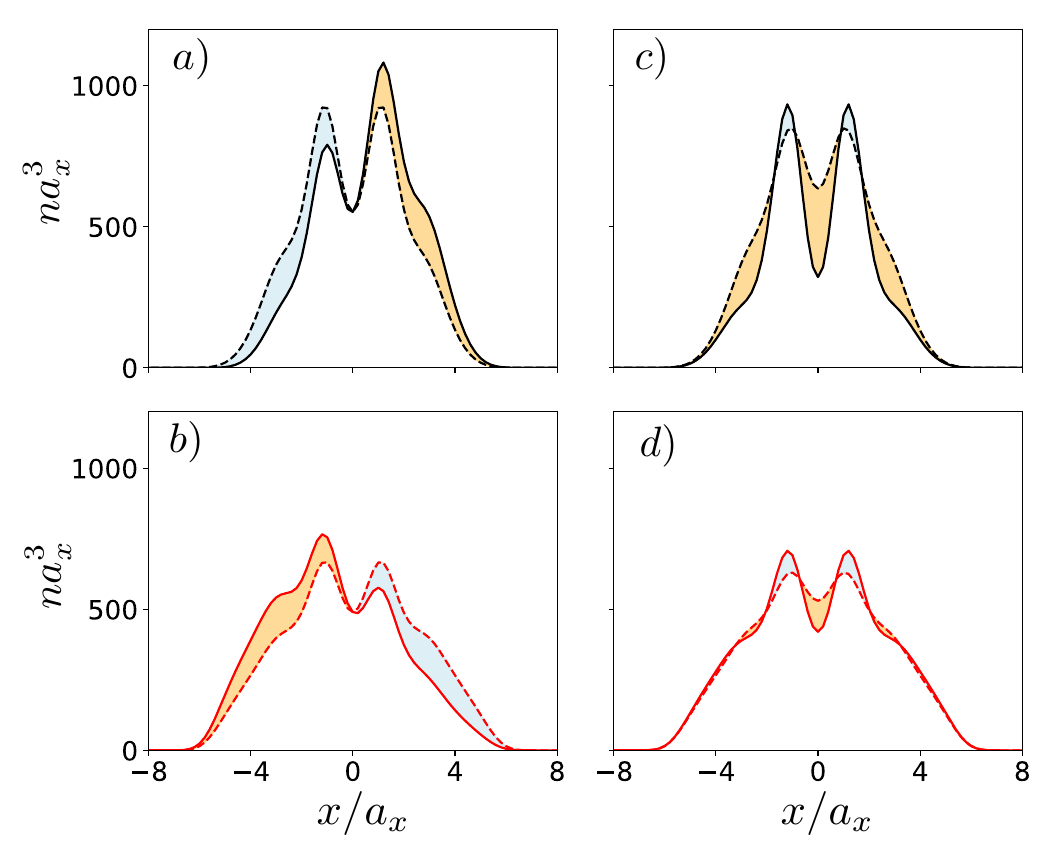}
    \caption{Ground-state density $|\psi_0^\sigma(\mathbf{r})|^2$~(dashed 
    lines) and ground-state density plus density fluctuations $|\psi_0^\sigma(\mathbf{r})|^2+\gamma\delta n_\sigma({\bf r})$, where $\gamma$ 
    is an arbitrary parameter that has been chosen to be large enough in order to 
    clearly distinguish the density fluctuations from the ground state density. Panels (a) and (b) depict the case of the supersolid Goldstone mode evaluated at $a_{12}=80\,a_0$, whereas Fig. c~(d) shows the Higgs mode obtained at $a_{12}=80.2\,a_0$. Top (bottom) panels correspond to component 1 (2). All other parameters are as in Fig.~\ref{fig:3}.}
    \label{fig:4}
\end{figure}


\subsection{Unmodulated-to-double-supersolid transition: Higgs and Goldstone modes}

Resembling the case of single-component dipolar condensates~\cite{Hertkorn2019}, at the transition the symmetric roton transforms into the Higgs~(amplitude) mode, which hardens abruptly in the supersolid regime, hybridizing with other modes. The pure Higgs mode, found at the bending depicted in detail in Figs. (a$_2$) and (b$_2$), is approximately a density mode, and $\delta n_1(\mathbf{r})$ and $\delta n_2(\mathbf{r})$ are oscillating in phase, see Figs.~\ref{fig:4}~(c) and~(d).



\begin{figure}[t!]
    \centering
    \includegraphics[width=0.9\linewidth]{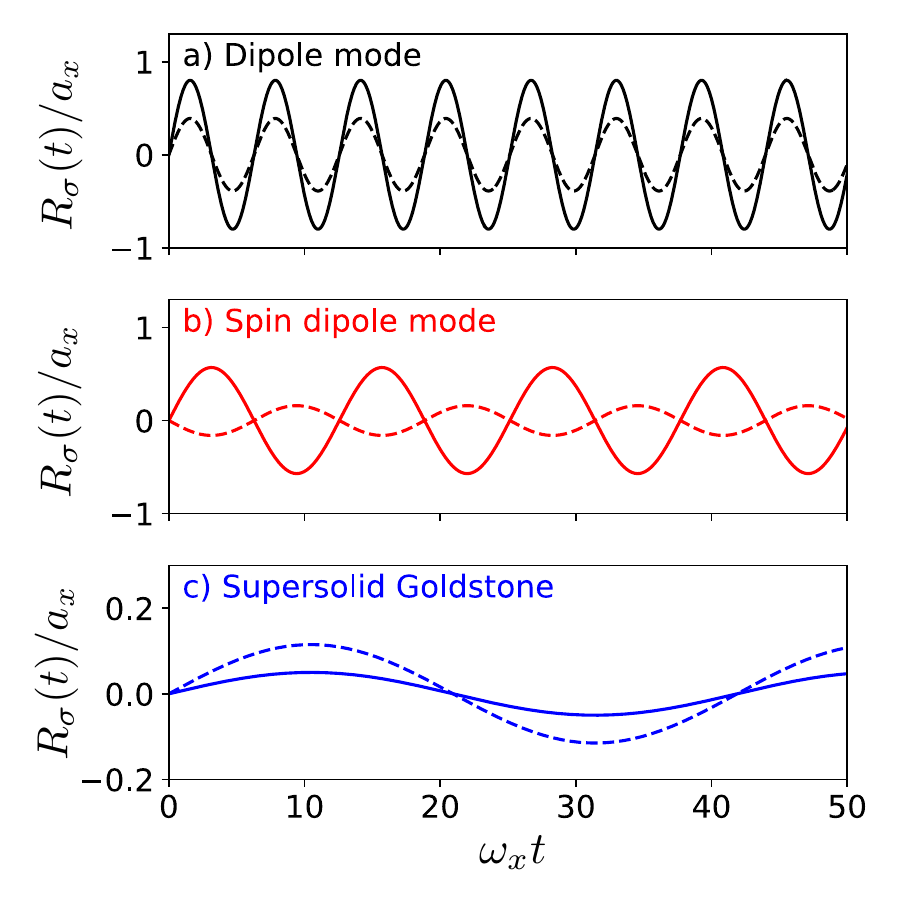}
    \caption{Oscillation of the center of mass of component 1~(solid) and 2~(dashed) for the dipole mode~(a), the spin dipole mode~(b), and the supersolid Goldstone mode~(c). The modes have been evaluated for $a_{12}=71\,a_0$. All other parameters are as in Fig.~\ref{fig:3}.}
    \label{fig:5}
\end{figure}

The three gapless Goldstone modes of the double supersolid in the infinite tube are mirrored in the trapped case into a clear three-mode structure of the low-lying discrete modes. The lowest-lying triplet of excitations comprises the evolved antisymmetric roton, and the two dipole modes that stem from the Goldstone modes of the unmodulated mixture. One dipole mode, with frequency $\omega_x$, is characterized by the in-phase oscillation of the centers of mass $R_\sigma(t)=[\int x\delta n_\sigma({\bf r})d{\bf r}]\sin(\omega t)$ of the two components, see Fig.~\ref{fig:5}~(a). The other dipole mode, presents an out-of-phase oscillation of $R_1$ and $R_2$, see Fig.~\ref{fig:5}~(b). The energy of this mode is maximal at the unmodulated-to-double-supersolid transition, decreasing into the double-supersolid regime, as seen in Figs.~\ref{fig:3}~(c) and~(d). Finally, the lowest-lying mode, which we denote as supersolid Goldstone mode, is characterized by a strong hybridization of density and spin. However, the mode has a strong weight in component 1, $P\simeq 1$. 
Since we may rewrite $S_\sigma = \frac{1}{2}\int d^3 r \theta_\sigma(\mathbf{r})\delta n_\sigma(\mathbf{r})$, the supersolid Goldstone mode exhibits significantly larger phase fluctuations in component 1 compared to component 2, due to the lower minimal density in component 1.
The supersolid Goldstone mode presents a much smaller amplitude of the motion of the centers of mass, see Fig.~\ref{fig:5}~(b). As for the case of a single-component condensate~\cite{Guo2019}, this occurs because the displacement of the crystalline structure is compensated by a particle imbalance in the opposite direction. Interestingly, although the center of mass of both components moves in the same direction with a small amplitude~(significantly more pronounced in component 2), the actual oscillation of the density imbalance occurs out-of-phase in the two components, see Figs.~\ref{fig:4}~(a) and~(b).



\begin{figure}[t!]
    \centering
 \includegraphics[width=0.9\linewidth]{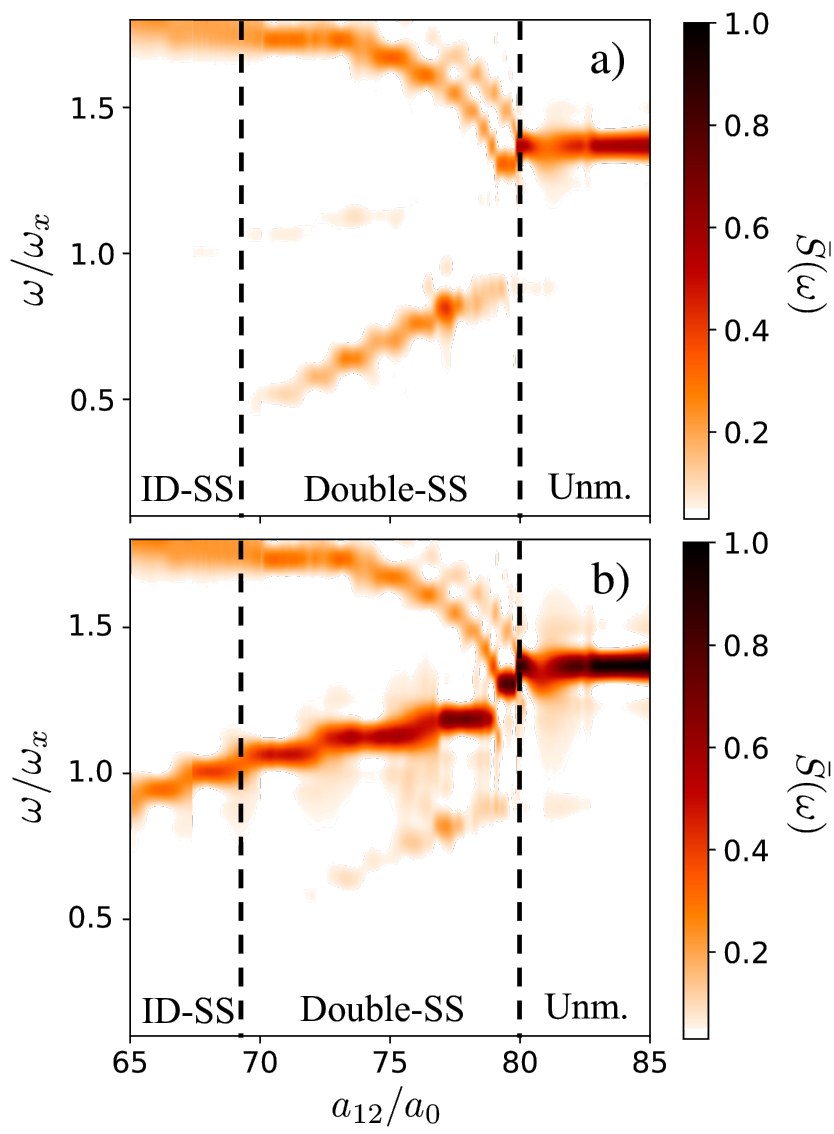}
    \caption{Strength $\bar{S}(\omega)$ of the signal of the axial breathing mode as 
    a function of $a_{12}$ for component $1$ 
    (panel a) and component $2$ (panel b). The results are normalized to the 
    maximum value between the two plots. All parameters are as in Fig.~\ref{fig:3}. Note that for each $a_{12}$ $\bar{S}(\omega)$ is normalized to the maximal value in any one of the two components.}
    \label{fig:6}
\end{figure}


\subsection{Axial breathing modes}

In contrast to the symmetric mixture, an asymmetric one allows probing of the double supersolid character by using a simple trap compression, which as mentioned above, induces the same $x^2$ perturbation in both components. The lowest breathing mode, which stems from the spin breathing mode of the unmodulated regime, acquires a marked spin-density hybridization at the transition, becoming almost fully dominated by the first component, $P\simeq 1$. As for the symmetric mixture, the density breathing mode of the unmodulated regime splits into two modes at the transition. While the hardening crystal mode retains a clear density character, the softening mode is also strongly hybridized and evolves into an almost pure mode of the second component, $P\simeq -1$. Hence, both softening superfluid modes couple with the compressional perturbation, resulting in a characteristic signal in the $\bar {S}(\omega)$ spectrum evaluated separately for each component, see Fig.~\ref{fig:6}. Whereas the hardening crystal mode is equally shared by both components, each superfluid mode is almost fully dominated by one of the two components, i.e. the superfluid breathing modes present a dominantly single-component nature. Note that the superfluid mode dominated by component 1 has a lower energy, as expected from the larger contrast, and hence, following Legget's upper-bound~\cite{Leggett1970}, a lower superfluid fraction. The compressional modes hence allow to probe not only the double supersolid nature, but also the markedly different superfluidity of both components.



\begin{figure}[t!]
    \centering
    \includegraphics[width=\linewidth]{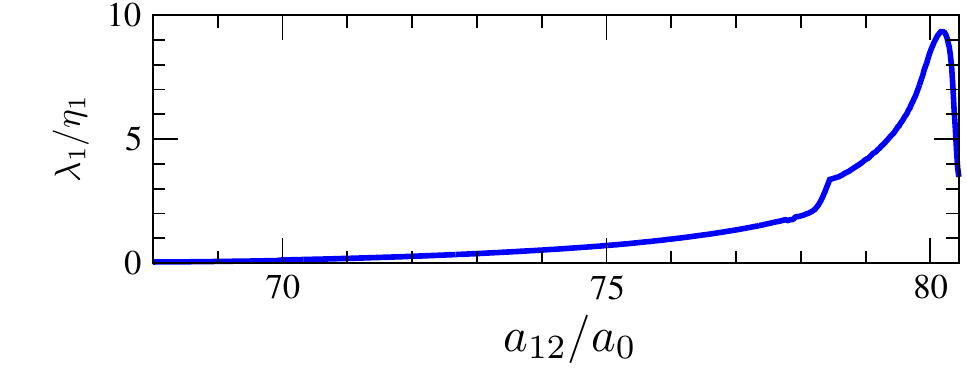}
    \caption{Ratio $\lambda_1/\eta_1$ as a function of $a_{12}$ obtained for the lowest-lying excitation mode. This ratio characterizes the phase in component 1, which when the ratio approaches zero only variates in between droplets, marking the onset of the ID regime for component 1. All parameters are as in Fig.~\ref{fig:3}. }
    \label{fig:7}
\end{figure}



\subsection{Transition into the ID-supersolid regime}

When $a_{12}$ is further reduced, the lowest-lying mode grows in energy down to $a_{12}\simeq 70\,a_0$, and then softens. This is due to the hybridization with a down-coming odd mode characterized by an out-of-phase oscillation of the two central droplets and the halo at the wings. At $a_{12}\simeq 70\,a_0$ the halo decouples from the two central droplets, resulting in a halo mode that gains energy abruptly. Note that, since the left and right part of the halo are disconnected, the symmetric and antisymmetric halo modes become energetically identical. As a result, the breathing mode associated with component 1 hardens abruptly, and disappears from the excitation spectrum $\bar{S}(\omega)$, see Fig.~\ref{fig:6}. The lowest-lying mode, which is a mode of the two central droplets, eventually softens at $a_{12}\simeq 65\,a_0$, marking the decoupling of the droplets in component 1. The vanishing coherence in component 1 is also evident from the monitoring of the phase fluctuations of that component. In particular, $\lambda_1/\eta_1$ approaches zero for $a_{12}\simeq 68\,a_0$, as shown in Fig.~\ref{fig:7}, indicating that the droplets present a constant phase, but are mutually incoherent. These results clearly mark the transition from the double supersolid into the ID-supersolid regime, which occurs for values of $a_{12}/a_0$ in very good agreement with the value obtained from the analysis of the density contrast~(Fig.~\ref{fig:1}).

In the ID-supersolid regime, the mixture is characterized by a spectrum with a clear two-mode structure. This mirrors the infinite-tube idealized scenario, where the ID-supersolid regime is characterized by two gapless Goldstone modes, associated with the crystalline modulation and the remaining superfluidity of component 2. Finally, for an even lower $a_{12}$ (not depicted), the system enters the ID-ID regime. The modes of component 2 perform a similar splitting into a halo mode (that abruptly hardens) and a droplet mode that softens down to zero energy. As a result, the low-lying spectrum acquires a simple structure characterized by purely crystal modes.

\section{Conclusions}
\label{sec:conclusions}

In this paper, we have studied the excitation spectrum of a trapped dipolar Bose mixture, and how this spectrum provides key insights about the nature of the double-supersolid phase, with two coexisting interacting superfluids with a generally different superfluid fraction. 
While we have focused on the simple, yet experimentally relevant, case of two-droplet supersolids due to numerical complexity, our conclusions are expected to hold more generally. 
In particular, the doubling of the superfluid breathing modes, and their basically single-component character, which should be relatively straightforward to monitor~\cite{Tanzi2019b, Natale2019}, would generally be a clear proof of the double-supersolid nature and the markedly different superfluid fraction of the two components in asymmetric mixtures. 
Moreover, the modes 
exhibit a highly nontrivial nature in what concerns their contribution to the modulation of the overall density, of the relative density~(the spin), and of the phases of the two components. 
This is in particular the case of the dipolar rotons,  the Higgs excitation, and the lowest-lying Goldstone modes, which present a very different density-spin hybridization. The different superfluid fraction of the two components becomes especially clear from the detailed analysis of the Goldstone modes, which reveal the crossover of one of the components, but not of the other, into the incoherent droplet regime.

\section*{Acknowledgements}

We acknowledge the support of the Deutsche Forschungsgemeinschaft (DFG, German Research Foundation) -- Project-ID 274200144 -- SFB 1227 DQ-mat within the projects B01 and A04, and under Germany's Excellence Strategy -- EXC-2123 Quantum-Frontiers -- 390837967).

\bibliography{bibliography.bib}

\end{document}